# AN ALGORITHM FOR DETECTING QUANTUM-GRAVITY PHOTON DISPERSION IN GAMMA-RAY BURSTS: DISCAN

Jeffrey D. Scargle[1], Jay P. Norris[2] and J. T. Bonnell[3]

[1] Space Science Division
NASA/Ames Research Center, Moffett Field, CA 94035-1000

[2] Denver Research Institute
University of Denver, Denver CO 80208

[3] Astroparticle Physics Laboratory
NASA/Goddard Space Flight Center, Greenbelt, MD 20771.

## ABSTRACT

DisCan is a new algorithm implementing photon dispersion cancellation in order to measure energy-dependent delays in variable sources. This method finds the amount of reversed dispersion that optimally cancels any actual dispersion present. It applies to any time- and energy-tagged photon data, and can avoid binning in both time and energy. The primary motivation here is the search for quantum gravity based dispersion in future gamma-ray burst data from the Gamma Ray Large Area Space Telescope (GLAST). Extrapolation of what is know about bursts at lower energies yields a reasonable prospect that photon dispersion effects consistent with some quantum gravity formalisms may be detected in sufficiently bright bursts. Short bursts have no or very small inherent lags, and are therefore better prospects than long ones, but even they suffer a systematic error due to pulse asymmetry that may yield an irreducible uncertainty. We note that data at energies higher than about 0.1 TeV may not be useful for detecting dispersion in GRBs. Of several variants of the proposed algorithm, one based on Shannon information is consistently somewhat superior to all of the others we investigated.

# 1. INTRODUCTION

The search for a physical theory uniformly valid in the large (strong field gravity, high energy physics, and cosmology) and the small (atomic and subatomic particles and the corresponding forces – strong, weak, and electro-weak) has yet to succeed. This failure is partly due to the lack of empirical tests relevant across these scales. A number of astronomical observations now under study may shed light on some generic theoretical concepts (Amelino-Camelia, et al. 1998, Biller et al. 1999; Schaefer 1999; Ellis et al. 2006; Martinez, Piran & Oren 2006). Here we describe a new algorithm to detect a possible dependence of the speed of light on photon energy in time series data with sharply defined structures but small photon counts. Suitable observations of gamma ray bursts at high (GeV) energies may soon be provided by the Gamma Ray Large Area Space Telescope[1], scheduled for launch in early 2008. There is considerable interest in the fact that such a *photon dispersion* effect may be observable at the higher photon energies that will be accessible to the GLAST instrument, serving as a possible test of some theories. See (Norris *et al*. 1999) for preliminaries.

In §2 we briefly review the current status of quantum gravity (the name given to this field, abbreviated QG) theories, describe photon dispersion and expectations for this effect in the sensitive energy range of GLAST's Large Area Telescope (LAT), namely 10 MeV to 300 GeV. In §3 is a discussion of relevant properties of bursts and their pulse structure, and the details of the important issue of extrapolation of what is known from previous lower energy observations to higher energies. Then §4 introduces DisCan, a data analysis procedure designed to detect and measure photon dispersion, and §5 shows the results of applying this procedure to synthetic data. The final section summarizes the prospects of using short GRBs for constraining energy-dependent time lags that may be consistent with some QG theories.

---

[1] GLAST (Atwood et al. 1994) is an international and multi-agency mission, consisting of two instruments, the Large Area Telescope (LAT) and the GLAST Burst Monitor (GBM). See http://www-glast.stanford.edu for a list of contributors and more details.

## 2. QUANTUM GRAVITY THEORIES

There is no consensus on the correct approach to the unification of small and large-scale physics. But it is frequently assumed that at very small scales space-time is quite different from its macroscopic appearance, and that the transition has something to do with the Planck time and spatial scales, described below. Ideas for the actual small-scale nature of space-time are quite diverse. They include a fuzziness (described quantitatively by a generalization of the Heisenberg uncertainty principle) called *quantum foam* by Wheeler (1982), and the completely opposite assumption of infinitely sharply defined sets of discrete points, *e.g.* as in causal set theory (Henson 2006). In turn, such lumpiness or discreteness may produce observable effects in the propagation of light, notably photon dispersion (the speed of light depending on the energy of the photon). Readers finding such views of space-time implausible should keep in mind that intuition based on ordinary human perception does not apply at even the atomic scale, much less the Planck scales yet more dramatically removed from the macroscopic world.

That quantum gravity effects should appear at the Planck scale is easily motivated using an argument embodied in the generalized uncertainty principle (Adler and Santiago, 1999). The numerical values of the Planck spatial and temporal scales are almost unimaginably small:

$$L_{Planck} = (G\hbar/c^3)^{1/2} \sim 1.6 \times 10^{-35} \text{ m} \qquad (1)$$

$$T_{Planck} = (G\hbar/c^5)^{1/2} \sim 0.54 \times 10^{-43} \text{ s} \qquad (2)$$

where G is Newton's gravitational constant, $\hbar$ is Planck's quantum constant, and c is the speed of light. From dimensional considerations alone, these combinations of physical constants are prime candidates for the scale of quantum gravity effects, because they bring together constants fundamental to gravity, quantum mechanics, and relativity. By including the effects of photons' gravity on measurements in the standard derivation of the Heisenberg uncertainty principle, Adler and Santiago show that there is a minimum uncertainty to the measured position of any entity, and to the time to which the measurement refers. These are the Planck scales given above.

A survey of quantum gravity publications reveals a surprising number of divergent approaches: string theory, non-commutative field theories, loop quantum gravity, causal set theory, doubly special relativity, Regge calculus, and many others. A very useful overview is provided by Mattingly (2005). None of these theories makes a definite prediction of the existence of photon dispersion.[2] However a simple argument, widely used to suggest a plausible energy dependence for the speed of light, postulates a perturbation of the ordinary dispersion relation expressible as a Taylor series.

---

[2] On the other hand there are negative results. Bombelli, Henson and Sorkin (2006) point out that a discrete space-time does not imply symmetry breaking (such as of Lorentz invariance) leading to modified dispersion relations. Hence causal set theory does not imply photon dispersion.

Following Ellis et al. (2002), if the resulting first-order coefficient ξ (expressed on the Planck scale) is positive and near unity, the difference in time of arrival for two photons with energy difference ΔE would be

$$\Delta t = t_H (\Delta E/E_p) \int h(z)^{-1} dz \quad (3)$$

where $h(z) = [\Omega_\Lambda + \Omega_M(1+z)^3]^{1/2}$, $E_p = (\hbar c^5/G)^{1/2} \sim 1.22 \times 10^{19}$ GeV is the Planck energy, and $\Omega_\Lambda$ and $\Omega_M$ are the usual cosmological and matter constants. Taking $H_0 = 71$ km s$^{-1}$ Mpc$^{-1}$ in a flat universe with $\Omega_\Lambda = 0.73$ and $\Omega_M = 0.27$, the Hubble time is $t_H = H_0^{-1} = 4.34 \times 10^{17}$ s. The time difference Δt corresponds to the difference in light travel time for photons differing in energy by ΔE traveling over the comoving distance $d_{CM}$. For ΔE = 1 GeV and $d_{CM}$ = 1 Gpc, Δt ≈ 8 ms. Such timescales are comparable to pulse widths in short GRBs. Moreover, the intrinsic spectral lags at the energies that have been well explored (by the Burst Alert Telescope, or BAT, of the *Swift* mission and the Burst and Transient Source Experiment, or BATSE, of the Compton Gamma Ray Observatory) are consistent with zero, with uncertainties of order a few ms for the brightest short bursts (Norris & Bonnell 2006). Yet, as with long bursts, short bursts' pulses tend to narrow at higher energies. Unlike long bursts, which can have as many as tens of pulses with considerable overlap, short bursts tend to have few pulses with little overlap. These four attributes – few narrow pulses with intrinsic peak times approximately independent of energy, but narrower at higher energy and little overlap – are particularly desirable for the purpose of exploring the possibility of quantum gravity photon dispersion.

The distance scale for short bursts is only qualitatively understood, with a handful of measured redshifts of host galaxies and other deduced redshifts (Berger et al. 2006). Ambiguities affect some individual measurements, when uncertainty in the afterglow position prevents definitive association with one host galaxy. Also, selection effects in the redshifts for short bursts may be more severe than for long bursts, given that their afterglows tend to be dimmer. For present purposes it is sufficient to note that, roughly speaking, $z \sim 0.1$–$1.0$ for short bursts, corresponding to $d_{CM} \sim 0.4$–$3.3$ Gpc, yielding $\Delta t \sim 3$–$28$ ms.

Long bursts detected by *Swift* would also appear to be good tools for dispersion searches, given that their redshifts tend to cluster in the range $z \sim 1$–$4$, corresponding to larger distances, $d_{CM} \sim 3.3$–$7.3$ Gpc. However, their median pulse widths and intrinsic spectral lags are 10–20 times longer (with tails to much longer timescales) than those in short bursts, and their pulses most often overlap (Norris 2002; Norris & Bonnell 2006). Thus, measurements which would provide clear characterizations of long burst pulses are made difficult by intrinsic spectral trends, and by pulses that are significantly wider than the maximum expected value for $\Delta t$ ($\sim 40$ ms at 5 Gpc under the above assumptions). In this discussion clean separation of pulses has been emphasized, not because separation as such is crucial, but because it entails large gradients in the profile, which enhance any algorithm's sensitivity to lags.

In conclusion, pulses in short GRBs, traversing Gpc distances and detected at GeV energies, are probably the best time markers available for detecting quantum gravity photon dispersion.

## 3. BURST MODELING AND SIMULATIONS

Our purpose here is to construct an energy-dependent model of a representative burst, in order to extrapolate to the higher energy gamma-ray regime. Varying values of the model parameters allows exploration of the constraints that can be placed on QG-based energy-dependent photon dispersion. Relevant parameters include burst intensity, duration, pulse width, pulse confusion, and *spectral lag*.[3] These studies indicate that relatively bright bursts will be required to place meaningful limits on photon dispersion effects. For this and additional reasons discussed below, we chose the very bright short burst GRB 051221a as the exemplar. An additional requirement is that the detection technology have smaller deadtime effects, such as those which rendered EGRET ineffective for precise timing of GRBs. (GLAST will nicely meet this requirement.)

### 3.1 *Modeling GRB 051221a at BAT Energies (15-150 keV)*

Soderberg et al. (2006) provide a complete description of GRB 051221a and its afterglow properties. This burst was instantaneously the brightest burst detected by the *Swift* mission to date, and greater than an order of magnitude more energetic in prompt γ rays (and in kinetic energy of the afterglow components) than all other short bursts for which redshifts are known. GRB 051221a lies in a dwarf star-forming galaxy at redshift $z = 0.5464$, thus at a comoving distance of ~ 2 Gpc.

Figure 1 shows the more intense portions of the burst at 1-ms resolution. The time profile consists of four strong, narrow pulses of widths ~ 5–20 ms and ~ five lower intensity pulses. The two most intense pulses reach ~ 175,000 counts s$^{-1}$, placing the burst in the upper ~ 3% of short bursts detectable by the BAT. The intense pulses are fortuitously separated by intervals comparable to or larger than the pulse widths, making modeling relatively easy. Based on a comparison with the much larger (nine year) BATSE sample, and accounting for the differences in spectral sensitivity compared to the BAT, such a bright short burst should be detected by *Swift* roughly once in 2-3 years.

---

[3] This term refers to energy dependent lags inherent in the gamma-ray burst emission mechanism, and/or radiative transfer in and around the burst, that has to be accounted for in any claim of detection of true quantum gravity effects.

Characteristic of short bursts, the average intrinsic spectral lags for the interval shown in Figure 1 are 0.0 ± 0.4 ms (0.8 ± 0.5 ms), between the 15–25 keV and 50–100 keV channels (25–50 keV and 100–350 keV), as reported in Norris et al. (2005a). Essentially, at BAT energies the pulse positions are nearly independent of energy. This qualitative difference in pulse behavior compared to long bursts may be attributable to the "curvature effect," with higher Lorentz factors obtaining for short bursts (Norris & Bonnell 2006; Aloy, Janka, & Muller 2005). The expectation then is that, due to the decreasing angle from which increasingly higher energy photons can reach the observer (Sari & Piran 1997 ; Zhang & Meszaros 2004), at LAT energies the pulses in short bursts will be even narrower, still exhibiting effectively no shift of pulse peak. How much narrower will remain a question until actual LAT observations of bursts.

To characterize the pulse shapes we use the same asymmetric pulse model employed to fit the pulses in long-lag bursts (Norris et al. 2005b):

$$I(t) \;=\; A\lambda \,/\, [\exp\{\tau_1/t\}\,\exp\{t/\tau_2\}] \;=\; A\,\lambda\,\exp\{-\tau_1/t - t/\tau_2\} \ \ \text{for } t>0, \quad (4)$$

where $\mu = (\tau_1/\tau_2)^{1/2}$ and $\lambda = \exp(2\mu)$. The peak intensity, $A$, occurs at $t = \tau_{peak} = (\tau_1\tau_2)^{1/2}$. The pulse width (measured between the two $1/e$ intensity points) and asymmetry are, respectively

$$w = \Delta\tau_{1/e} = \tau_2 (1 + 4\mu)^{1/2}, \tag{5}$$

$$\kappa = (\tau_{decay} - \tau_{rise}) / (\tau_{decay} + \tau_{rise}) = (1 + 4\mu)^{-1/2} = \tau_2 / w. \tag{6}$$

The pulse fitting procedure, including steps taken to constrain parameters where pulses overlap substantially, is described in Norris et al. (2005b).

Figure 1 illustrates the total fit (dark line), and individual pulse fits (gray lines), with the background model ($\approx 8.3$ counts ms$^{-1}$) subtracted. The values for the fitted pulse parameters are listed in Table 1. In lieu of the formally fitted time of pulse onset, $t_s$ (which sometimes occurs at an intensity several orders of magnitude below the peak intensity, in which case $t_s$ is not indicative of the visually apparent onset time) we tabulate $t_{eff}$, defined as the time when the pulse reaches 0.01 times the peak intensity. Near the peaks of the two most intense pulses the residuals are relatively high, but the fit is adequate for our central purpose of obtaining a parametric representation of the pulse shapes, which can be extrapolated to the high-energy gamma-ray regime, as well as mutated to produce bursts of different intensity, pulse width, etc.

## 3.2 *Extrapolation to LAT Energies*

To produce simulated bursts in the energy regime where the LAT will be sensitive, the model fit described above needs to be extrapolated more than three orders of magnitude, from ~ 300–1000 keV to > 1 GeV. We are nearly completely ignorant of GRB pulse behavior over most of this range, and so some assumptions must be made.

For long bursts, the dependence of pulse width on energy was found to follow a power-law form at BATSE energies,

$$w \propto E^{-\nu}, \tag{7}$$

with $\nu \sim 0.33$–$0.43$, depending upon the sample details and specific measurement procedure (Fenimore et al. 1995; Norris et al. 1996). A similar dependence appears to obtain for short bursts as well (Norris et al., in preparation). To cover the reasonable possibilities, we made synthetic bursts with a range of values for $\nu$: 1/3, 1/4 and 1/5, where the last value effects little additional pulse narrowing above BATSE energies. The degree of pulse asymmetry is assumed to remain constant. From Eqs. (5) and (6) we see that, for the transformation $\{\tau_1 \to \lambda\tau_1, \tau_2 \to \lambda\tau_2\}$, $\tau_{peak}$ and $w$ increase by $\lambda$, whereas the asymmetry $\kappa$ is invariant. We also simulated the pulse peak times to be independent of energy.

Power-law spectra with indices ranging from 1.6 to 2.4 are indicated for bursts detected by EGRET (Dingus, 1995), and in particular the short burst GRB 930131 (Sommer et al. 1994; Kouveliotou et al. 1994). To extrapolate the fluence reported for GRB 051221a by Golenetskii et al. (2005) to high-energy gamma rays, we assumed a spectral index of 2, yielding $\sim 0.025$ photons cm$^{-2}$ incident above 30 MeV.

For GRB 051221a, at a comoving distance of 2 Gpc, the quantum gravity dispersion as estimated above would be $\tau_{QG} \approx 17$ ms / GeV. For our idealized synthetic bursts we chose a round value for convenience, $\tau_{QG} = 20$ ms / GeV. This dispersion was applied to the synthetic photons, delaying them linearly as a function of energy – thereby introducing a simulated extrinsic lag of the opposite sign that is usually seen in long bursts. In the resulting synthetic LAT time profiles, on average a 1 GeV photon lags a 30 MeV photon by almost 20 ms. The effect would be perfectly recoverable were it not for the intrinsic widths and separations of pulses, which are comparable to the introduced extrinsic effect at BAT energies, and narrower at LAT energies by an unknown degree.

For each value of the power-law index ($\nu$) for pulse width, 1000 bursts were prepared, with randomized photon energies and times. These simulated burst data streams, representing the flux incident on the LAT instrument, were then fed to a simple computer routine approximating the LAT detection characteristics and energy response.

For our purposes the relevant instrumental characteristics are the energy-dependent effective collecting area, and the energy dispersion. The LAT has an effective area profile, $A_{eff}(E)$, with threshold ~ 10–30 MeV, rising to half maximum near 120 MeV, and continuing to rise above 1 GeV, to ~ $10^4$ cm$^2$. We adopted the function $A_{eff}(E)$ available on-line from the GLAST LAT Collaboration[4]. This function describes the estimated $A_{eff}(E)$ on the instrument axis. The bursts were simulated to arrive at an angle of 45° to the axis, reducing $A_{eff}$ by the factor ~ 0.7. The average number of photons detected was 98 ± 10. In addition, normally distributed variance was added to the true photon energies to simulate the detection process. Above 100 MeV, we set $\sigma_E/E = 0.10$, increasing linearly to $\sigma_E/E = 0.20$ as 100→30 MeV.

---

[4] http://www-glast.slac.stanford.edu/software/IS/glast_lat_performance.htm

### 3.3 *Simulated LAT Bursts*

Figure 2a plots the times and energies of each photon in a synthetic burst with $\nu = 1/3$. The pulses are sufficiently narrower than the pulse separations such that there is no possibility of assignment of a photon to an incorrect pulse. Figure 2b shows a different realization, now with $\nu = 1/5$. Assignment of photons to wrong pulses is now possible. Notice that over the full BATSE energy range (~ 25–1000 keV) pulses would narrow by a factor of ~ 0.3. An additional narrowing factor of ~ 0.3 over the range 1 MeV to 1 GeV, as obtains for $\nu = 1/5$, seems feasible. Thus Figures 2a and 2b may bracket the appearance of a burst like GRB 051221a at LAT energies. Notice that the number of photons with energies > 1 GeV is ~ 5.

Figure 2c is the same realization as shown in Figure 2a, but with application of an energy-dependent dispersion of 20 ms / GeV. This dispersion coefficient multiplies the *incident* photon energy, since this is what governs any propagation effects. However, in applying the dispersion cancellation methodology discussed in the next section, only the *detected* energies are known. In Figure 2 the detected energies are plotted on the ordinates. At the lower energies in Figure 2c, the photons have been displaced only slightly compared to Figure 2a, whereas above ~ 100 MeV, the delays become evident. In particular, the highest energy photon ($E_{inc}$ = 14.3 GeV, $E_{det}$ = 13.9 GeV), belonging in the third pulse, is displaced 0.286 s from t ≈ 0.18 (Figure 2a) to t ≈ 0.46 (Figure 2c), past the last pulse. Nevertheless, even with 20 ms / GeV dispersion, it is evident by visual inspection – for a burst as fluent as our extrapolation of GRB 051221a – which pulse each photon originally belonged too, except for the few highest energy ones.

In summary, two important effects would prevent a near perfect measurement of the dispersion coefficient: All pulses have finite width, which as we describe below, will result an irreducible uncertainty incurred in any dispersion estimation algorithm. We addressed our ignorance of GRB pulses at high energies by simulating pulse widths with a plausible range of energy dependences, parameterized by ν. Also, the dispersion cancellation procedure must make use of the measured photon energy, whereas the hypothetical QG effect is dependent upon the true energy. We found that use of detected, rather than true, energies increased uncertainties in the recovered dispersion parameter by ~ factor of 2. Another possible source of error is discretization of photon time tags, which for the LAT may introduce < 10 μs absolute errors in timing. We discretized the photon times to 10 μs quanta in order to estimate the magnitude of this effect, which turns out to be negligible compared to the effects of finite pulse width and energy measurement uncertainty.

For the correct parameter value the transform will restore the correct relationships between the photon times, and therefore the time profile will be maximally sharp. The following sections discuss several appropriate measures of sharpness or complexity.

## 4. DATA ANALYSIS METHODS

This section describes a new method for analysis of time-and-energy-tagged data to detect, or place an upper limit on, photon dispersion. As far as we know, all other methods require binning of the raw photon data in both time and energy. Typically time is binned preparatory to a cross-correlation or wavelet analysis, and energy is binned very coarsely -- most frequently into two bins (e.g., Ellis et al. 2006). Replacing data points with their counts in a set of pre-defined bins results in two deleterious effects. First, a lower limit is set on the resolution that can be obtained, namely the width of the bins. Second, the results of any subsequent analysis are dependent on the choices of bin sizes and locations. Often these choices, and therefore the resulting limitations, are arbitrary and not set by astrophysical or instrumental considerations. In addition, binning of a very small number of events is problematic in a fundamental way.

Furthermore, the misconception that bin counts must be large in order to be useful ("good statistics") makes things even worse by encouraging adoption of large bins. The primary advance reported here is an algorithm that can be entirely bin-free, thus avoiding the information losses implied by the issues just discussed.

To use the algorithm one must have time- and energy-tagged photon data, plus a model of the dispersion effect. The model is simply a formula giving the time delay as a function of photon energy. In this formula will be a parameter specifying the magnitude of the effect. (Depending on the details of the physical theory, there could be more than one parameter, but the application would be the same.) For a given value of this parameter the algorithm shifts each photon's measured arrival time by the negative of the time shift specified by the model. Then it finds the optimal parameter value by examining the time profile (or "light curve") of the data transformed in this manner for a set of parameter values. Optimal in what sense? Since the true shape of the time profile is not known, a direct model fit is not possible.

Instead, note that for incorrect parameter values the relative photon times will be scrambled randomly (due to the distribution of photon energies), blurring the time profile. For the correct parameter value the transform will restore the correct relationships between the photon times and the time profile will be maximally sharp. The problem is to maximize a measure of sharpness or complexity, several of which are described in §4.3.

An important aspect of this method is a complete separation of the way photon energies and photon times are incorporated. The energies determine the time shifts meant to undo the putative dispersion, and the times (actually the shifted times) only enter the time series representation.

## 4.1 The Quantum Gravity Model

The first step is to specify a parametric model for the dispersion. For now we model only the quantum gravity time delay as a function of energy, ignoring energy-dependent delays inherent to the GRB, foreground events from nearby sources, and other effects. The most general physical relation between the observed and true arrival times — absent dependence on other radiation properties, such as polarization — is

$$t_i^{obs} = f(t_i^{true}, E_i^{true}, \theta) , \qquad (8)$$

where f is an arbitrary function and $\theta$ represents one or more parameters. By true arrival time we mean the time at which the photon would have arrived had it suffered no quantum gravity time shift:

$$t_i^{true} = g(t_i^{obs}, E_i^{true}, \theta) \qquad (9)$$

for some function g, the inverse of f. Our method operates on the observed times, photon by photon, with the transformation in equation (9). This operation gives a set of putative arrival times (as a function of the model parameters), and is meant to undo, or cancel, the physical delay represented by equation (8). That is to say, the goal is to make the transformed photon arrival times equal to the actual emission times plus the constant delay due to the overall transit time.

Most authors have assumed a linear delay, specializing equation (8) to

$$t_i^{obs} = t_i^{true} + \theta E_i^{true}, \qquad (10)$$

with a scalar delay coefficient $\theta$, in units of time per unit energy. Throughout we report results in terms of this coefficient,[5] in milliseconds per GeV, a convenient unit if the Planck scale governs as discussed in §2. The actual physical process need not be of this linear form, but for present purposes, and comparison with others' work, we henceforth use this model. (However, note that the method is quite general and applies without essential modification to any model for the dispersion.) Accordingly the transformed photon arrival times are given by

$$t_i' = t_i^{obs} - \theta E_i^{obs}. \qquad (11)$$

This shift would exactly cancel the actual lag, except that $E_i^{obs} \neq E_i^{true}$, as discussed near the end of §3.3.

---

[5] Reporting time delay averaged across two fiducial energy bands makes the results more difficult to interpret, for example because of dependence on the spectrum of the source.

## 4.2 Representing the Time Series

For all of the cost functions but one (that based on photon time intervals), the next step is to construct a representation of the photon rate as a function of time (i.e. the burst time profile), from the arrival times transformed as described in the previous section. Three approaches described here are based on bins, cells, and blocks. The latter two are effectively bin-free and based in part on previous work (Scargle 1998; Scargle, Norris and Jackson 2007). No doubt other representations are possible. Note that photon energy is taken into account, photon by photon, in the same way for all representation methods, by transforming arrival times as described in the previous section.

The first, counts in equally spaced time bins, is easily implemented directly:

$$x_n = \text{ number of photons in bin n / size of bin n} . \qquad (12)$$

The number of bins, their sizes and locations free parameters; determining their best values is difficult. An empirical method for selecting the number of bins is described in §5 below. See Kevin Knuth's code repository[6] for an interesting procedure for selecting bin size.

The second approach is even simpler than binning: assign to each photon an interval dt, defined as average of the distances to that photon's two closest neighbors (the one just prior to it and the one just after it).[7] Denoting the width of this *photon cell* by $dt_i$, a rough estimate of the intensity (rate in photons per unit time) near photon i is just

$$x_i = 1 / dt_i. \tag{13}$$

---

[6] http://knuthlab.rit.albany.edu/optBINS.html

As seen in Figure 3 this cell-based representation, while somewhat choppy in appearance, effectively captures the salient features of the variation. The main benefit of this representation is that it completely avoids the information losses of binning described above. In effect it has the ability to reveal variation on arbitrarily small time scales, not limited by choice of a predefined bin size. For the parameter estimation task at hand, the main shortcoming (choppiness) is largely ameliorated by averaging over many photons.

The computation of these cells only requires the first difference of the ordered array of photon arrival times. If an interval is zero the above formula breaks down. This problem is easily solved by assigning two (or more) photons to the corresponding cells. In practice it is useful to do so whenever multiple photons have the same, or nearly the same, arrival times. Thus a slightly more general form of equation (13) is actually used, namely:

$$x_n = \text{ number of photons in cell n / size of cell n.} \tag{14}$$

Except for the case of clusters of identical or very close times, the numerator here is 1, giving the same result as in Eq. (13). Eq. (14) seems identical to Eq. (12), but the hidden difference is that the bins in the latter, typically all of the same largish size, are replaced with the smallish and unequal cells in the former. Moreover, the single cell representation is data-adaptive, and its bins are really cells whose sizes and locations are determined by the data, not by the data analyst. Hence Eq. (14) can be thought of as a bin-free version of Eq. (12).

---

[7] This interval consists of all times closer to that photon than to any other photon; i.e., it is the cell of a one-dimensional Voronoi tessellation (Scargle, Norris and Jackson 2006) of the times.

The third representation, based on the Bayesian Blocks algorithm (Scargle 1998, Scargle, Norris and Jackson 2006), is also bin-free. The procedure starts with the cells above, and considers partitions of the observation time interval consisting of contiguous blocks of the cells. As described in detail in the references, the algorithm yields the optimal such partition – that is, the piece-wise constant, or step function model, that best fits the data. Here, as with the raw cell representation of the profile, no pre-selected bins are involved. For details of how a set of intensity values can be computed from time-tagged photon data, the reader may consult the references given above.

The cost function based on average intra-pulse photon intervals, described below in section 4.3.3, works with the individual photon times directly, and does not require an explicit time series representation.

## 4.3   Measuring Sharpness: The Cost Function

The next step is to quantify the sharpness of time profiles constructed, say, by one of the methods described in the previous section. Examples of concepts that might form the basis of such a measure include information content, entropy, complexity, total variation -- or others such as fuzziness, opposite to sharpness. This *cost function*[8] is to be evaluated as a function of θ from the values of the time series representation, here denoted $x_n$.

We have experimented with several measures, including Shannon, Renyi and Fisher information, total variation, self-entropy, and minimum average intra-pulse photon interval. For the dispersion-recovery procedures we performed a grid of 800 trials over the range of θ from 0 to 40 ms/GeV – to bracket the amount added to the simulated bursts as described in §3.3.3 – in steps of Δθ = 0.05 ms/GeV. The results for several cases are described in Section 5.

### *4.3.1 Information content: Shannon and Renyi*

Shannon defined a quantity to measure the amount of information conveyed by specifying a value of a discrete variable, in terms of the probability distribution of that variable (Shannon & Weaver 1949; Cover & Thomas 1991). Since smoothing or blurring a time profile that has fine detail degrades its information content, any measure of information is a candidate for our cost function. The estimate of θ is the value that maximizes the information content of the time profile derived from the transformed photon arrival times in Eq. (11).

---

[8] In different disciplines and in the context of fitting models to data, many different terms are used for the quantity to be optimized, including *fitness function, risk, residuals, error measure, objective function*, etc. Here we use for any sharpness measure to be maximized, expressed as a function of the dispersion parameter(s) θ, the generic term *cost function*.

Start with any of the three time series representations discussed above, that is

$$x_n = \text{counts of photons in } \{\}_n / \text{size of } \{\}_n, \qquad (15)$$

where $\{\}$ stands for one of the three choices bin, cell, or block. Then form the corresponding normalized probability distribution

$$p_n = x_n / \Sigma x_n. \qquad (16)$$

The information measures are then

$$I(\text{Shannon}) = \Sigma p_n \log(p_n), \text{ and} \qquad (17)$$

$$I(\text{Renyi}) = -(1-\alpha)^{-1} \log [\Sigma p_n^\alpha], \qquad (18)$$

where all sums are over all bins, cells, or blocks, $n = 1, \ldots, N$. Renyi information contains a parameter $\alpha$; for $\alpha = 1$ Renyi information is not defined, but it can be shown that

$$\lim_{\alpha \to 1} I(\text{Renyi}) = I(\text{Shannon}) \qquad (19)$$

With opposite signs, these formulas may be familiar as defining the corresponding *entropy*, since information and entropy are negatives of each other (Brillouin 1962).

*4.3.2 Total Variation and Fisher Information*

Both Shannon and Renyi information measures are invariant to an arbitrary permutation of the bins, as they are sums of a quantity defined for each bin with no reference to bin order in general, or to adjacent bins in particular. In contrast, total variation and Fisher information take into account possible correlations between neighboring bins, cells or blocks. Frieden (1998) refers to this at the local nature of Fisher information. The formula for total variation takes bin-to-bin differences explicitly into account:

$$\text{Total Variation} = \Sigma \, | \, p_n - p_{n+1} \, | \, . \tag{20}$$

Fisher information is often defined in terms of derivatives of the distribution with respect to its parameter, starting from the *score*

$$V = \partial \log( p(x; \theta) ) / \partial \theta ) \, . \tag{21}$$

In turn Fisher information is the variance of the score:

$$I \, (\text{Fisher}) = E \, [ \, \partial \log( p(x; \theta) ) / \partial \theta \, ]^{\,2} \tag{22}$$

(Cover and Thomas 1991). A more convenient form, especially for time series applications like that considered here, is (Frieden 1998):

$$I \, (\text{Fisher}) = \Sigma \, [ \, p_n^{\,½} - p_{n+1}^{\,½} \, ]^{\,2} \, . \tag{23}$$

As remarked by Frieden, this form of Fisher information "simply measures the gradient content …" of the time series. Hence both measures, being sensitive to more of the information content of the data, are potentially superior to, say Shannon and Renyi information. In practice we have found that the total variation and Fisher information measures behave very similar to each other and are not significantly superior to the other information measures. We believe this is due to that fact that the first-difference operation (part of both Fisher and total variation measures) amplifies noise in the data, making the resulting cost function relatively noisy.[9]

### 4.3.3 *Average intra-pulse interval*

GRB time profiles commonly consist of multiple pulses. The final measure discussed here is motivated by consideration of such pulse structure, and in particular the statistics of the intervals between successive photons inside and outside of pulses. Note that, when the dispersion cancellation is optimal, the photons in a given pulse will be contained within the original (undispersed) interval covered by the pulse; hence the intra-pulse photon intervals will on average be small. Also, by definition the photon rate is higher near the pulse peak, so the intra-pulse intervals will tend to be smallest there. Furthermore, since pulses are narrower at higher energy, the higher energy photons are nearer to the peak. Hence, from several points of view higher energy photons are more sensitive indicators of dispersion. The most fortuitous arrangement is when the intervals between pulses are comparable to or greater than pulse widths.

In fact, the above reasoning is also implicit in the usefulness of the other measures of dispersion recovery, which have formal bases in statistical theory and were chosen for their appropriateness to measure entropy, sharpness, or complexity. This leads us to define an *ad hoc* measure explicitly designed to consider only the photon intervals *within* pulses. One such measure is the minimum of the averages of the intra-pulse photon intervals

---

[9] Indeed, the performance of these cost functions substantially improves if the profile is subject to a procedure that diminishes large amplitude spikes in the data. To do this we simply replaced any amplitude above some threshold with that threshold (upper hard truncation.

$$C(\theta) = \text{Min} \{ \Sigma \Delta t_i / N_{intra} \} \qquad (24)$$

where the $\Delta t_i$ are the $N_{intra}$ intervals between photons within pulses, such that $N_{intra} = N_{tot} - N_{inter}$, $N_{tot}$ is the total number of photon intervals, and $N_{inter}$ is the number of intervals between pulses. $N_{inter}$ can be estimated using, e.g. the subset of photons with energies > 200 MeV, by equating the number of significant pulses to the number of local maxima in a Bayesian block representation of the time profile. The value of $\theta$ which gives the minimum average intra-pulse interval is then considered the best dispersion coefficient for the burst.

## 5. PERFORMANCE COMPARISON OF COST FUNCTIONS

This section describes the application of the methods discussed in §4 to gamma-ray burst photon times and energies simulated as described in §3. We have explored various time series representations (bins, cells, and blocks) combined with various cost functions (Shannon, Renyi and Fisher information measures, total variation, and photon interval distributions). In applications to other contexts, the analyst should explore these and other relevant representations, combined with a variety of cost functions. The differences among the behaviors on our test data are relatively minor, and do not warrant an exhaustive exposition. Rather this section concentrates on the more significant differences between results for various types of simulated data, as will be explained below.

But first a few remarks on the binned representation described in §4.2. Its use is straightforward if the pre-binned data are all that is available. Starting from raw unbinned data is more challenging because it is desirable to select an optimal number of bins. For the burst model with $\nu = 1/3$ the panels in Figure 4 show, for the four indicated cost functions, the dependence of the estimated dispersion parameter $\theta$ on the number of bins. The mean of the optimal $\theta$ is plotted against the number of bins, with the scatter indicated with vertical bars; both quantities are averaged over the random burst realizations. The optimal value is that which yields results closest to the known value of $\theta$ inserted into the data. Note that for all four cost functions there is moderately large error for small numbers of bins, but for more than a dozen or so bins the value levels out close to the correct $\theta = 20$ ms/GeV. The variance decreases somewhat with increasing number of bins. The errors and variances for Fisher information and total variation are larger than for Shannon and Renyi information. The main conclusions of this figure are that the optimum model is more or less insensitive to the number of bins, as long as this number is greater than or equal to a reasonable value (here on the order of 10-20), and that the estimate is significantly biased toward smaller values of $\theta$ over the range covered. Note that over the range shown there is no minimum acceptable bin size, even in the limit of one point per bin. In practice it may be useful to reduce the variance by averaging over a range of bin sizes. The corresponding plots for the other two GRB profiles are similar, although the errors and standard deviations are progressively larger as the pulse overlap increases.

Figures 5 shows plots of various cost functions, for the binned time series representation, as a function of $\theta$. Note the well-defined maxima for all five of the cost functions, very near the correct value $\theta = 20$ ms/GeV. The figure legend gives the actual values of $\theta$ at which the maxima occur (and vertical lines below the curves indicates the locations also). Figure 6 and 7, the same plots for the cell-based and Bayesian block representations respectively, exhibit more scatter but in most cases have maxima near the correct value. (Fisher information and total variation for the block case being the most significant exceptions.) The results shown here are for the maxima of the curves evaluated at θ at evenly spaced intervals of 0.05; improvement by controlling the larger variances could be obtained by careful fitting of curves, such as low-order polynomials, to the data points.

Naturally the accuracy of recovery of the dispersion coefficient θ depends on burst characteristics: intensity, pulse overlap, pulse width, and spectral lag. To explore some aspects of these dependences we varied the model of GRB 051221a developed in §3 to realize six additional models – three short bursts, and three long bursts. For two short burst models the overall intensity was increased and decreased a factor of 3, resulting in sets of realizations with ~ 300 and 30 photons, respectively. For the third case, the most fluent pulse, #8, was moved to nearly coincide with the second most fluent pulse, #4 (see Figure 1). We positioned pulse #8 so that at lower LAT energies the pulse overlap for the width parameter ν = 1/3 was negligible; however for ν = 1/5, the pulses overlap to the extent that only one peak is present below ~ 300 MeV.

For the three long burst models, the pulse widths and peak times were stretched by a factor of ten, leaving the pulse asymmetries unchanged. Since the intensity was held constant, the number of photons in a model burst increases by a factor of ten to ~ 1000. Two models were constructed at this intensity, one with no intrinsic lag – as in the short burst models – and one with energy-dependent lags. Thus the resultant lag per photon was varied proportionally to pulse width, as $w \propto E^{-\nu}$, [see Eq. (7)]. For the present experiment, we only simulated the shift of the pulse peak – which constitutes about half of the total observed spectral lag at BATSE energies – with the average value being ~ half the full rise time of a pulse (Norris et al. 1996). Within a given burst lags vary among the pulses, so in this model we also varied the pulse lags randomly ± 20% from pulse to pulse, drawing from a uniform distribution. A third long burst model was constructed with the intensity decreased a factor of ten, to yield a comparable number of photons (~ 100) as in the standard short burst modeled after GRB 051221a. Each experiment was performed for the three values of the pulse width parameter, ν, 1000 realization per value.

Figures 8-10 illustrate the accuracy in the estimated dispersion parameter for three representative cost functions: Shannon information, total variation, and minimum average photon interval, respectively. For each of the seven models shown, short vertical lines denote the median $\theta_{opt}$ and thick (thin) horizontal lines indicate the 68% (90%) coverage interval, as determined from the 1000 realizations. These intervals are quantiles directly computed from the data, not just multiples of the empirical variance. The top four and bottom three cases are short and long burst models, respectively. It is clear that across all cases and cost functions, as pulse overlap (moderated by the parameter ν, depicted by the triads of coverage intervals) increases, estimation accuracy declines.

Several other trends in recovery accuracies for the different cost functions and model cases are apparent. First, for the short bursts Shannon information performs marginally better than the intervals measure; and for the intervals measure the coverage intervals are always smaller than those for total variation. On average for the first three short burst cases, the 68% errors for Shannon, intervals, and total variation are ~ 1.00, 1.15 and 1.40 ms/GeV, respectively.

The intervals measure out-performs Shannon only for case 4, the low-intensity short burst. For the long burst cases, Shannon information is always better, with 68% errors ~ 1.4× and 2× smaller than those for the intervals measure and total variation, respectively. Also, the intervals cost function appears to be biased toward lower and lower values of $\theta_{opt}$ as pulse overlap increases; the other two cost functions do not appear to suffer this trend.

Second, intensity matters. The long burst case (#5) where the short burst model is stretched a factor of ten, resulting in ~ 1000 photons per simulation, has uncertainties only ~ 20% higher for $\nu = 1/3$; however for $\nu = 1/5$, the uncertainties range from 1.6 to 2.0 times the standard short burst case (#2). The central values are comparable, irrespective of $\nu$. The conclusion is that increased fluence helps compensate for wider pulses.

Third, however, for a fixed number of photons, short bursts do better. Comparing the standard short burst with long burst case #7 (e.g., in Figure 8, Shannon information), the central values are similar, but uncertainties are 3-5 time larger, with the larger uncertainties pertaining as $\nu$ decreases and pulses widen. Thus the efficacy of narrow pulses in short bursts is apparent.

When spectral pulse lag is introduced in the long burst (case #6), the dispersion estimation procedure must deal with an intrinsic and extrinsic effect without a priori knowledge of dependences on energy. Here, the scale of the introduced intrinsic lag was made comparable to the pulse rise time to peak, and similar in magnitude to the extrinsic effect. The central values are about four times more discrepant from the true value than for the standard short burst (~1.5 ms/GeV vs. 0.3-0.4 ms/GeV, for Shannon information).

The obvious is proven: intrinsic spectral lag introduces an important effect that is difficult to separate from any extrinsic effect, especially at energies where the low fluences will offer few chances to characterize pulse shapes.

One noticeable systematic error is that the estimated θ values are consistently somewhat less than $\theta_{true}$ = 20 ms/GeV (by ~ 2%, *e.g.*, for Shannon information). Simulations verified that this systematic error is due to pulse asymmetry, and thus represents a possible irreducible uncertainty, given that pulses will be sparsely represented at GeV energies, and therefore pulse shape may not be accurately extrapolated from lower energies.

We performed several experiments, not detailed here, attempting to characterize the accuracy of dispersion recovery at lower energies, ~ 10 keV to 30 MeV, where the GLAST/GBM is sensitive. The conclusion is that, if QG temporal dispersion effects appear near the Planck energy scale, a detector with an effective area > 1000 $cm^2$ sensitive to at least 300 MeV would be necessary to probe a dispersion of ~ 20 ms/GeV/Gpc. Even then, our simulations show that the uncertainties would be ~ 100% for a source distance of 5 Gpc.

Paradoxically, photons of even higher energy (approximately a few tenths of a TeV and higher) may not be useful without better relative accuracy of the energy measurement. For if the uncertainty ΔE/E is of the same order as at lower energies, the error in the effective time delay, being proportional to ΔE, will exceed the GRB pulse separation, thwarting the measurement of the delay coefficient. For example, ΔE/E = 10% at 1 TeV translates into an uncertainty in $\varepsilon_{\tau QG}$ on the order of 1s at 1 Gpc, much larger than temporal structures in a typical GRB. Thus, countervailing the increased lever arm at higher energies is an error incurred by any dispersion estimation technique which utilizes transformed photon times in the way described for DisCan.

## 6. SUMMARY AND CONCLUSIONS

A brief discussion of theoretical considerations underlying the hope that timing measurements of gamma-ray bursts may detect dispersion effects that, in turn, could allow fundamental tests of the physics of photon propagation through space-time, was given in §2. All theoretical approaches allow the existence of energy-dependent photon delays, with the possible exception of causal set theory. No theoretical approach makes an unambiguous prediction for dispersion.

The enthusiasm for the empirical side of this topic seems to stem mostly from the assumption that quantum gravity shows itself in the form of higher order terms in a Taylor series expansion of the classical dispersion relation. While lacking a specific physical basis, this idea has the benefit of being fairly generic and independent of the conceptual framework of the quantum gravity theory. Under the further assumption that the Planck scale sets the order of magnitude of effect, careful timing of data on gamma-ray burst profiles may indeed allow detection of such delays. Our view is that this topic should be approached in an exploratory frame of mind. Any observational results, either detections or upper limits, will have to be evaluated carefully before fundamental conclusions can be reached. Of course this cautiousness is further necessitated by the possibility of inherent energy-dependent emission times, as well as other problems. The importance of the potential physical understanding that could result from one or more observable test of quantum gravity theories means that the observations are eminently worth pursuing, motivating the data analysis considerations in the rest of this paper.

We needed a body of data for developing and testing algorithms for measuring relative time delays, and §3 details the procedures that generated synthetic data of known properties and with known energy dependent delays. An essential feature of the time profiles is that they consist of a number of pulses, and a sequence of 3 cases corresponds to different pulse widths, and therefore different degrees of overlap in time between successive pulses.

The key innovation of this paper is the development and testing of a class of algorithms for estimating energy-dependent time lags (of the sort that might be consistent with quantum gravity theories), from a data stream in which individual photon arrival times and energies are measured. The algorithm consists of these steps:

   1) adopt a model of the time delay
   2) restore the arrival time of each photon according to the model
   3) construct the time profile using these transformed times
   4) find the parameter optimizing a measure of this profile
      sharpness

(Items 3) and 4) for the cost function based on intra-pulse photon intervals are somewhat different.)

The examples in this paper are all for a linear model in step 1), but the method is easily extended to any model, linear or nonlinear, with one or more parameters. Step 2) simply transforms the arrival times to what they would have been; it recovers the true value for the correct model and parameter values, except for an instrumental dispersion. We consider three modes of representation of the profile in step 3), in terms of bins, cells, or blocks. The latter two methods are in effect unbinned in time, therefore not discarding information by artificially degrading the time resolution. We also consider 5 different measures of the sharpness of the profile, including quantities based on information (negative entropy), variance, and intra-photon intervals. The performance for various combinations of profile representation, sharpness measure, and nature of the burst, are depicted in Figs. 4-10. Information measures of the Shannon and Renyi type are quite robust, and do not degrade significantly with pulse overlap. Fisher information and the total variation metric potentially make use of more information, since they are sensitive to the first difference of the profile; but an offsetting factor is the noise amplification that accompanies this differencing. The other cost functions, based on variance and inter-photon time intervals, have comparable performance in our simple test cases. If one had to select one cost function based on the studies depicted here, it would be that based on Shannon information. However, analysts are encouraged to explore other cost functions for specific contexts.

The studies conducted here are not a complete exploration, but do indicate that lag vs. energy coefficients can be measured with an accuracy on the order of 1 millisecond per Gev of photon energy difference, using the kind of gamma-ray burst data that will by produced by GLAST or other future high-energy astronomical observatories. While there are uncertainties in the extrapolation of knowledge of GRB behavior to high energies, our simulation studies indicate that the intrinsic properties of bright, short gamma ray bursts probably make them most suitable for such observations.

This effectiveness is mainly because their intrinsic lags are expected to be small. Future studies of spectral evolution of GRBs in general may suggest inclusion of long bursts. In any case measurements of a number of bursts at different distances will be helpful in disentangling intrinsic and quantum gravity lags (Ellis et al., 2004, 2006).

We are grateful to Joe Bredekamp and the NASA Applied Information Systems Research Program (AISRP) for support, Rafael Sorkin, the GPB theory group (including Ron Adler, Frances Everett, Bob Wagoner, Alex Silbergleit, and James Overduin), and Michael Peskin for helpful comments. We thank the anonymous referee for suggesting a number of improvements to this paper. We would like to acknowledge use of the GRB data, publicly available at the *Swift* website, http://swift.gsfc.nasa.gov/docs/swift/swiftsc.html.

TABLE 1. Pulse Fit Parameters

| A | $t_{eff}$ | $\tau_1$ | $\tau_2$ | $\tau_{peak}$ | $w$ | $\kappa$ |
|---|---|---|---|---|---|---|
| 18.8 | 111 | 5.64E+03 | 4.24E+00 | 156 | 51 | 0.083 |
| 19.3 | 172 | 2.45E+03 | 2.60E+00 | 198 | 29 | 0.090 |
| 62.4 | 204 | 8.57E+02 | 6.44E+00 | 239 | 44 | 0.146 |
| 141.6 | 256 | 3.79E+02 | 3.64E+00 | 274 | 24 | 0.155 |
| 105.7 | 290 | 2.55E+00 | 1.43E+01 | 295 | 23 | 0.610 |
| 120.3 | 314 | 3.42E+00 | 1.28E+01 | 320 | 22 | 0.571 |
| 16.7 | 332 | 2.77E+03 | 5.30E+00 | 375 | 51 | 0.104 |
| 102.3 | 394 | 6.31E+00 | 2.12E+01 | 405 | 38 | 0.561 |
| 15.3 | 460 | 4.65E+03 | 3.58E+00 | 499 | 43 | 0.083 |

Notes. All temporal parameters are in milliseconds, and amplitudes are in counts ms$^{-1}$. Pulse asymmetry [$\kappa$, see Eq. (7)] is dimensionless.

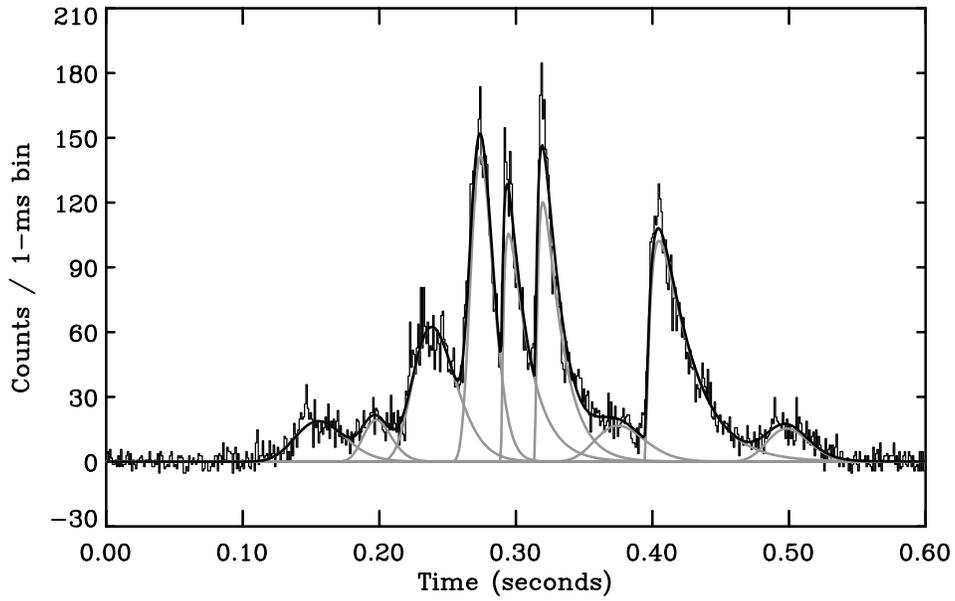

Fig. 1. – Time profile (count in 1-ms bins vs. time in seconds) of GRB 051221a in the energy range 15–350 keV, illustrating the prompt hard X-ray emission as detected by *Swift's* BAT. The dashed line represents an essentially flat background fit to the approximately 25 second intervals directly preceding and following the time interval shown.

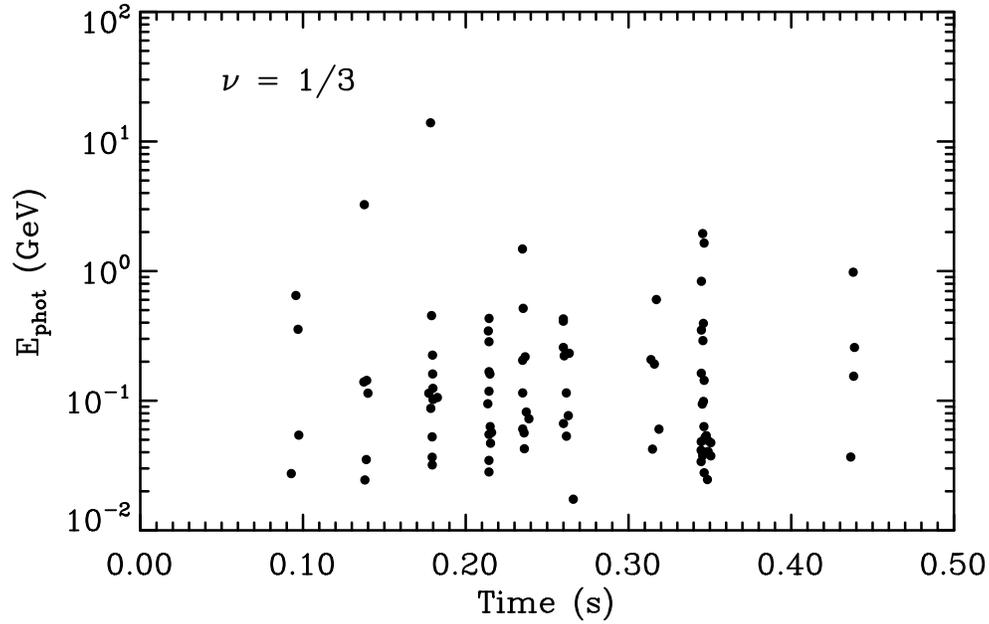

Fig. 2a. – Energy vs. time for a sample of photons simulated from the high-energy extrapolation described in the text, for $\nu = 1/3$, and with no time lag imposed.

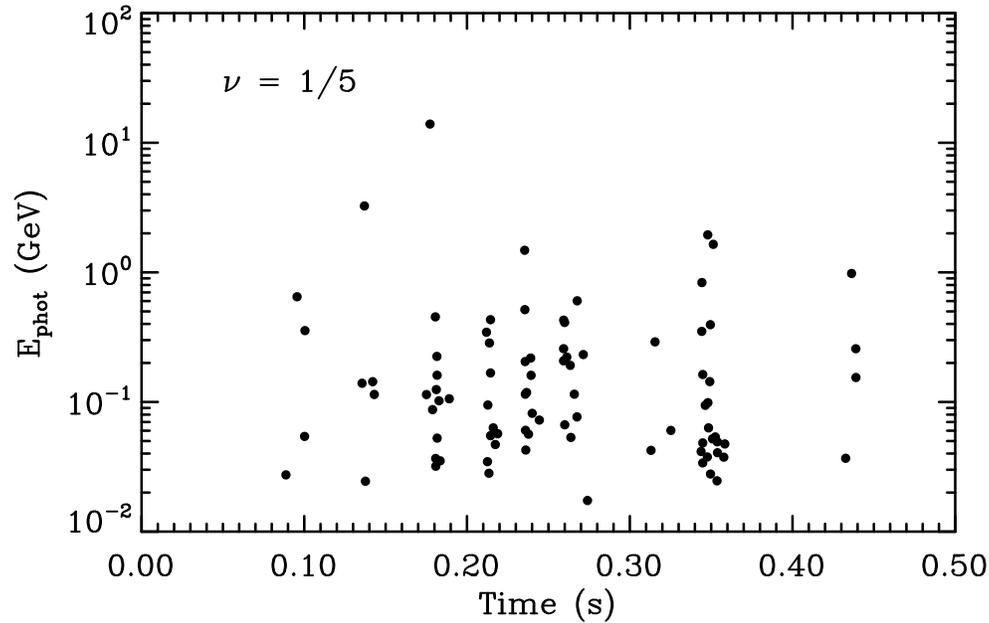

Fig. 2b. – For comparison with Fig. 2a, this plot is for $\nu = 1/5$, for which pulse narrowing at high energy is less pronounced.

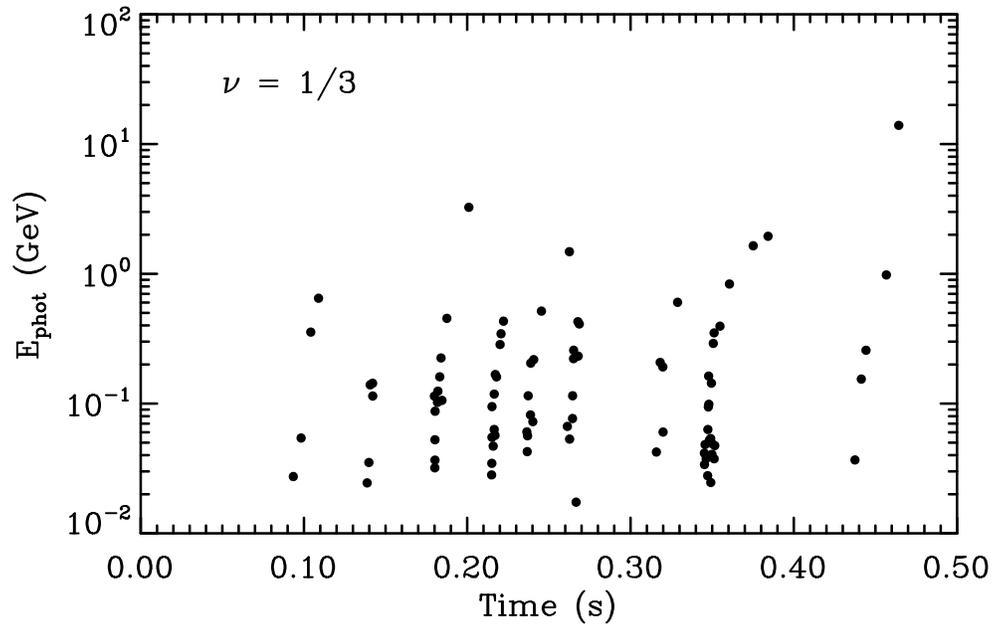

Fig. 2c. – This is the same as Fig. 2a, but a delay is applied to each photon according to its energy, using a coefficient of 20 ms/GeV. The curvature of the set of events in individual pulses is easily seen.

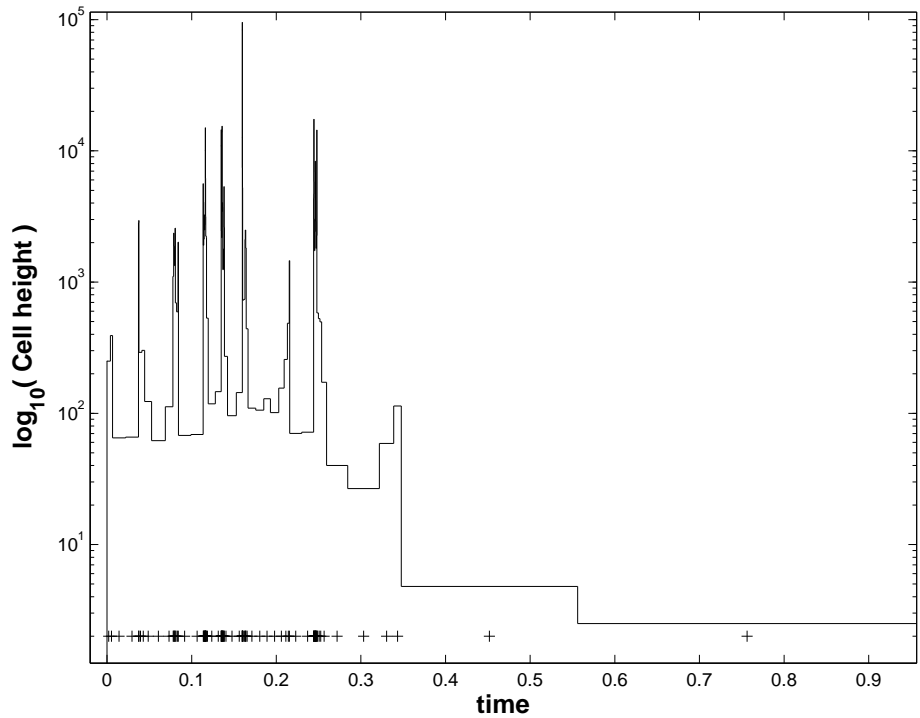

Fig. 3. – Sample cell representation of the 98 events in a simulated GRB observation. Plotted for each photon (shown as a "+" symbol at the bottom of the plot) is a rectangle of height equal to the reciprocal of the average time interval to that photon's two nearest neighbors. Thus blocks with larger heights indicate higher count rates. This representation is useful in itself, albeit a bit noisy, and as well serves as input for the Bayesian blocks routine.

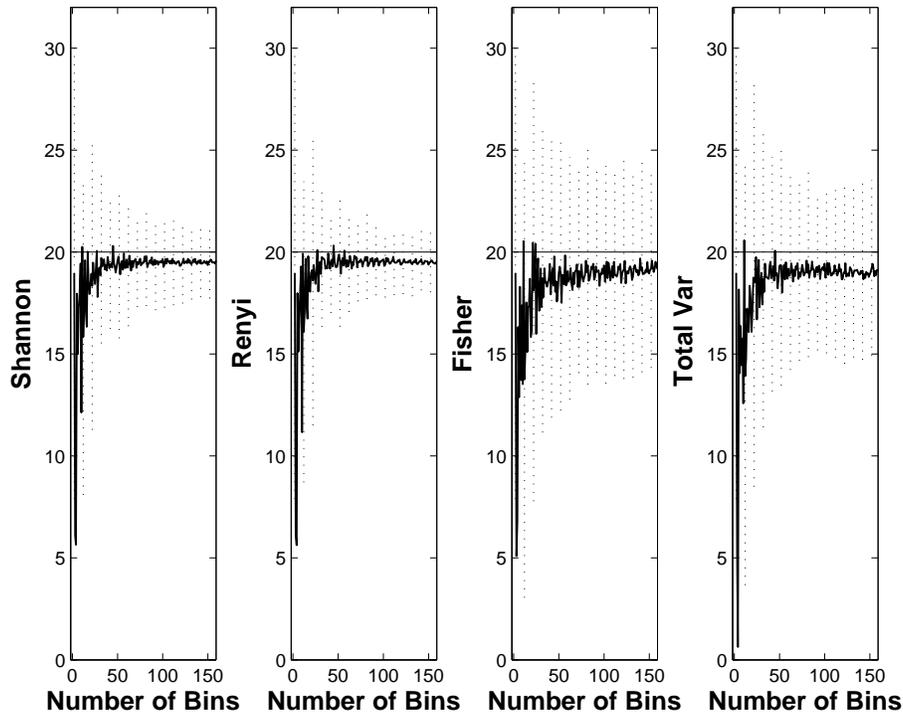

Fig. 4. – This figure demonstrates the effect of the number of bins on the estimated value of θ, for four measures of profile sharpness. The optimal value of θ (averaged over 1000 realizations) is plotted against the number of bins used in the computation (all values between 2 and 160; for reference, 100 bins corresponds to a time interval of 35 ms). The true value (20 ms/GeV) is shown as a horizontal line. Representative error bars (dotted lines) are the corresponding ensemble standard deviation (1σ).

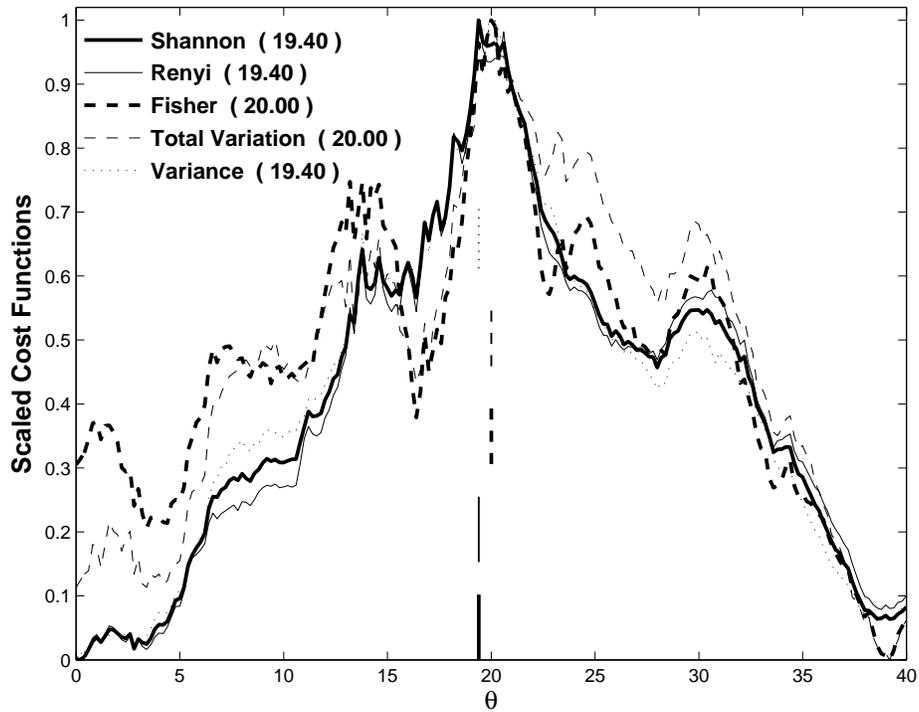

Fig. 5. – Five measures of sharpness of the binned profile representation, as a function of the model parameter θ, for a single realization of a simulated burst with $\nu = 1/3$. The optimal values of θ (locations of the maxima) given in the legend, and also plotted as vertical lines, are to be compared to the true value of 20 ms/GeV.

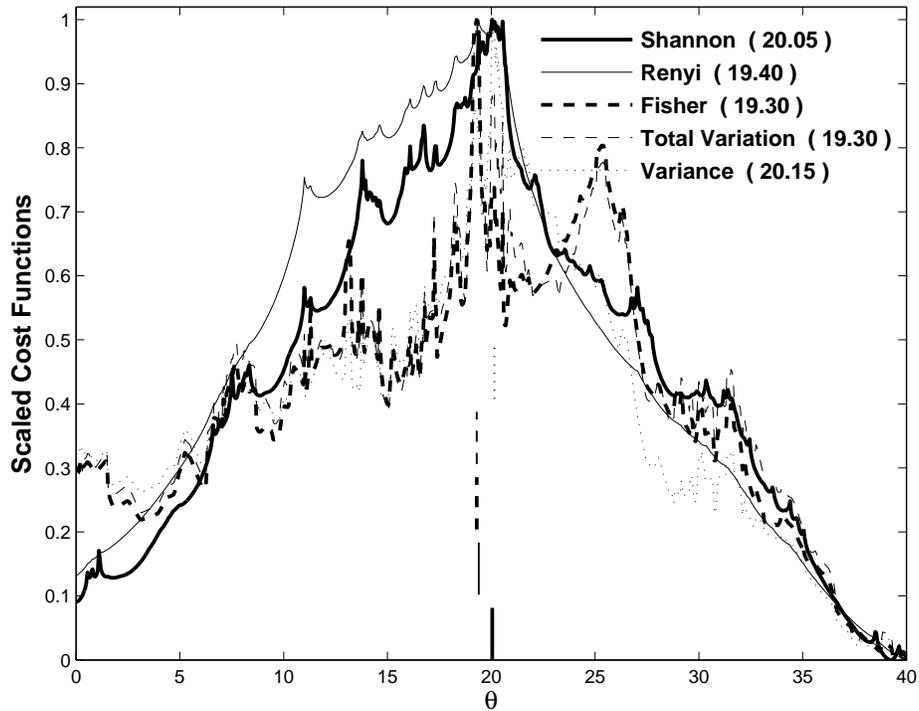

Fig. 6. – Similar to Figure 5, but for the cell representation of the time series. This and the following figure show a somewhat spikier dependence than the previous one, presumably because of the smoothing effect of the binning in Fig. 5.

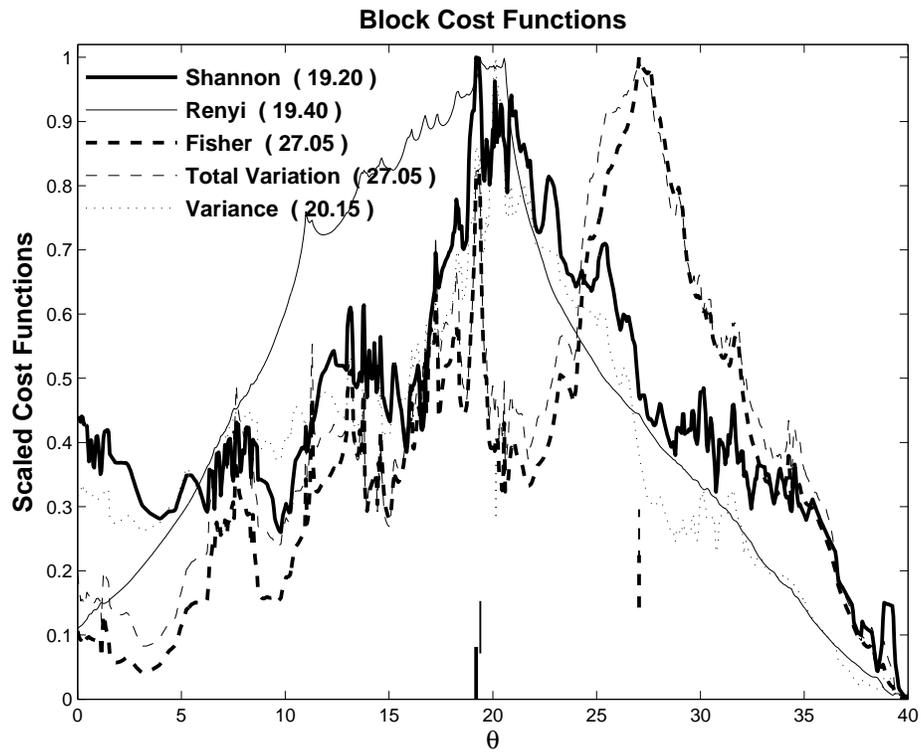

Fig. 7. – Similar to Figure 5, but for the Bayesian block representation.

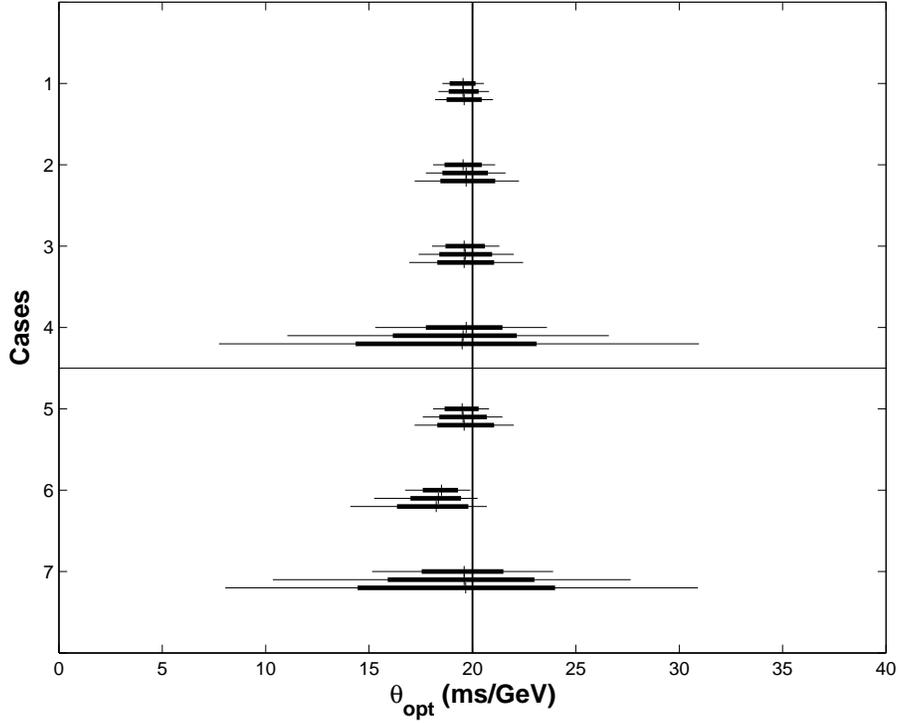

Fig. 8. – Values of θ obtained estimated using the Shannon information measure (Eq. 17) applied to the cell representation of 1000 simulated time series. Small tick marks are the median estimated dispersion (true value: 20 ms/Gev), and the thick and thin horizontal lines correspond to the ±68% (±90%) coverage intervals, respectively. From top to bottom the four short burst cases are: (1) intensity × 3, (2) intensity × 1, (3) intensity × 1 with pulses #4 and #8 adjacent as described in the main text, and (4) intensity × 1/3. The three long burst cases at the bottom are: (5) intensity × 1, (6) intensity × 1 with intrinsic energy-dependent spectral lags, and (7) intensity × 0.1. Within each of these 7 models, three values of the pulse width parameter (decreasing from top to bottom: ν =1/3, 1/4, and 1/5) are depicted.

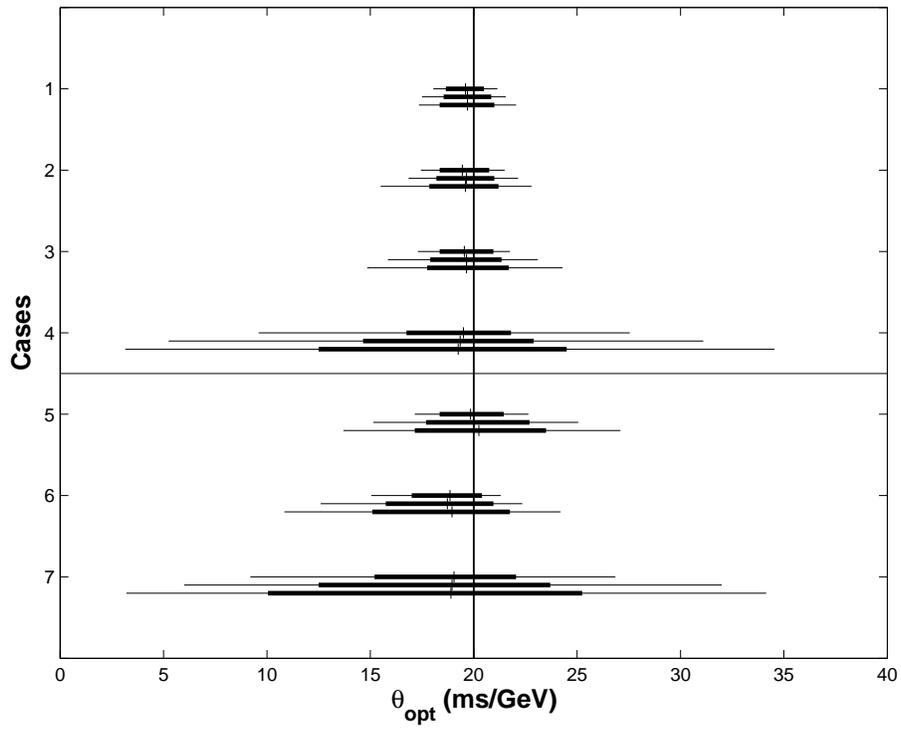

Fig. 9. – As in Figure 8, for the estimates obtained with the total variation measure (Eq. 20) applied to the cell representation.

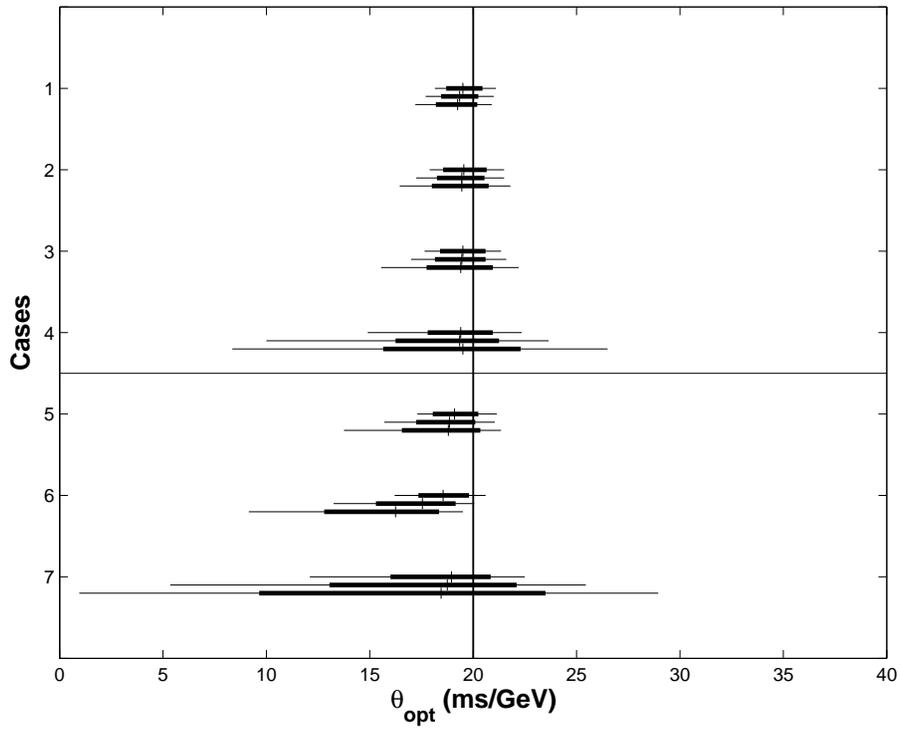

Fig. 10. – As in Figure 8, for the estimates obtained with the minimum average interval measure (Eq. 24) applied directly to the event data.